\documentclass{ws-rv9x6}
\usepackage{subfigure}     
\usepackage{ws-rv-van}     
\makeindex
\begin{document}

\chapter{Superfluid Pairing in Neutrons and Cold Atoms}\label{ra_ch1}

\author[J. Carlson, S. Gandolfi, and A. Gezerlis]{J. Carlson, S. Gandolfi}
\address{Theoretical Division, Los Alamos National Laboratory, Los Alamos, New Mexico, 87545, USA}

\author[J. Carlson, S. Gandolfi, and A. Gezerlis]{Alexandros Gezerlis}
\address{Department of Physics, University of Washington, Seattle, Washington, 98195, USA }

\begin{abstract}
Ultracold atomic gases and low-density neutron matter are unique in that they exhibit pairing
gaps comparable to the Fermi energy which in this sense are the largest in 
the laboratory and in nature, respectively. This strong pairing regime, or the crossover between
BCS and BEC regimes, requires non-perturbative treatments.  We describe Quantum Monte Carlo
results useful to understand the properties of these systems, including infinite homogeneous matter
and trapped inhomogeneous gases. 
\end{abstract}

\body

\section{From cold atoms to neutron stars and back}
\label{section:INTRO_general}

Although the energy and momentum scales of cold atomic gases and
atomic nuclei differ by many orders of magnitude, we can gain insight
into superfluid pairing in the strongly correlated regime by comparing
and contrasting the two systems. Dilute neutron matter and ultracold
atomic gases near infinite scattering length (unitarity) are similar in that
their equation of state and pairing gaps, when measured in terms of the
Fermi energy, are comparable.
For this reason these systems can be viewed as
``high-temperature superfluids'', even though one occurs at very small and the
other at very large temperature. Since the pairing gap $\Delta$ 
is roughly proportional to the
critical temperature $T_c$, both these systems exhibit very strong pairing, the
strongest ever observed before in nature or experimented with in the
laboratory. In cold fermionic atoms the
particle-particle interactions can be tuned experimentally, thus also mimicking the setting of
low-density neutron matter, which is beyond direct experimental reach.
More specifically, at low densities $s$-wave scattering can be described using only two parameters: 
the scattering length $a$ and the 
effective range $r_e$. 

\subsection{Ultracold Atomic Gases}
\label{section:INTRO_general_CA}

For ultracold fermionic atoms
at low temperature, 
\cite{Giorgini:2008} 
the superfluid phase arises across the entire spectrum
of weak to strong attractive interactions. 
Experiments with $^6$Li or $^{40}$K have an interparticle spacing 
that is significantly larger than the 
characteristic length of the interaction.
Over the last decade experimentalists managed to vary the 
interaction strength (the scattering length $a$)
across a resonance, through a regime known as ``unitarity''.
The cold atom gases are very dilute, so the typical scale of the interaction range
(or the effective range $r_e$) is much smaller than the average interparticle spacing. 
For these broad Feshbach resonances we have $k_F r_e \ll 1$, where $k_F$ is the Fermi wave vector
($\rho = k_F^3 / (3\pi^2)$), so at fixed density
the effective-range can be taken to be very small, essentially zero.
It may be possible to use narrow and wide resonances in cold atoms to study
the case of varying $r_e$ experimentally, \cite{Marcelis:2008} and thus directly 
simulate neutron matter.

A large variety of equilibrium and dynamic properties have
been measured in cold Fermi atom experiments. Experiments using $^6$Li at
Duke University \cite{Luo:2009} and at ENS \cite{Navon:2010} have measured the ground-state
energy of the system, essentially finding it to be in good qualitative agreement with Quantum
Monte Carlo (QMC) predictions
\cite{Carlson:2003,Astrakharchik:2004,Carlson:2005,Gezerlis:2008,Forbes:2011} 
The ground-state energy per particle is conventionally given in
units of the energy of a free Fermi gas at the same density $E_{FG} \equiv 3
E_F / 5 = 3 \hbar^2 k_F^2 / (10m)$ as $E = \xi E_{FG}$. 
Recent sign-free auxiliary field QMC calculations \cite{Carlson:2011} and experiments \cite{Zwierlein:2011}
give very precise results $\xi$ at unitary, $\xi = 0.372(5)$ and $0.375(5)$, respectively.

The great advantage of cold atom systems is that they offer quantitative
experimental results in the strong-coupling regime with simple interactions.
An important example is the pairing gap at unitarity. Experiments at MIT and Rice
probed lithium gases with population imbalance (also called ``polarized''
gases). An MIT
experiment \cite{Shin:2008} established the phase diagram of a polarized gas,
revealing spatial discontinuities in the spin polarization. This experiment
was then used \cite{Carlson:2008} to extract the pairing gap, which was found
to be approximately half of the Fermi energy $E_F$, in good agreement with QMC
calculations \cite{Carlson:2005}. The  gap is conventionally given in units of 
$E_F$ as $\Delta = \eta E_F$. The MIT group later used RF spectroscopy to independently determine
the gap, finding it to be in agreement with the afore-mentioned calculation
and extraction. \cite{Schirotzek:2008}

\subsection{Neutron matter and neutron drops}
\label{section:INTRO_general_NM}

As already mentioned, at very low densities, neutron matter is very similar to cold Fermi atoms.
The neutron-neutron scattering length is fixed and large ($a \approx -18$ fm),
but by varying the density we can probe different values of $k_F a$.
Neutrons in the inner crust of a neutron star are expected to pair in the $^1\mbox{S}_0$
channel, \cite{Lombardo:2001,Dean:2003}
at higher densities, also important in neutron stars, the effective range becomes important as do the
repulsive parts of the neutron-neutron interaction, and higher partial waves.

The inner crust of a neutron star contains a neutron gas embedded in a sea
of ions.  The properties are largely determined by the EOS and pairing
of homogeneous matter. Pairing in neutron matter has been studied
for many decades, leading to a large spread
of predictions of the $^1\mbox{S}_0$ pairing gap even for this idealized system.
The gradient terms in the density functional
and the behavior of the pairing gap in an inhomogeneous system
are potentially important for the crust of a neutron star
and can be studied simulating neutrons in external fields~\cite{Gandolfi:2011}.

It may be possible to access superfluidity in neutron star matter observationally:
superfluidity in a neutron star is
often used to explain its dynamical and thermal evolution, impacting
the specific heat, bremsstrahlung, and pair breaking/formation \cite{Page:2009, Page:2012}.
Additionally, a cooling mechanism that makes use of
superfluid phonons \cite{Aguilera:2009, Cirigliano:2011} has been proposed.
Whether  this mechanism is competitive
to the heat conduction by electrons in magnetized neutron stars or not is a
question that is directly correlated to the size of the gap.

Neutron matter 
and neutron drop computations also hold 
significance in the context of traditional nuclear physics:
equation of state results at densities close to the nuclear saturation density
have been used for some time to constrain  Skyrme and other density functional approaches 
to heavy nuclei, while the density-dependence 
of the $^1\mbox{S}_0$ gap in low-density neutron matter 
has recently also been used to constrain Skyrme-Hartree-Fock-Bogoliubov treatments 
in their description of neutron-rich nuclei \cite{Chamel:2008}.
Recent {\it ab initio} results for neutron drops point to 
a need for more repulsive gradient terms in inhomogeneous neutron matter, and a reduced 
isovector spin-orbit and pairing strength compared to standard functionals \cite{Gandolfi:2011}.

\section{BCS and Quantum Monte Carlo methods}

BCS theory has been critical to understanding many of the pairing properties of nuclei.
Quantum Monte Carlo many-body simulations, on the other hand, 
have been used for some time to
calculate the equation of state of strongly-correlated systems, e.g. liquid helium. In such systems,
however, QMC methods were unable to reliably calculate pairing gaps because of the
vast difference in scale between the energy of the entire system (of the order
of eV's per particle) and the pairing gap (of the order of meV). Thus, 
the same feature of strongly paired fermionic
systems (namely the large pairing gap) that precludes the application of mean-field
theories is precisely the reason that allows many-body simulation techniques
to be used. 

\subsection{Weak coupling}
\label{sec:exact} 

We first briefly review the weak-coupling regime where exact results are
available. At extremely low densities ($| k_F a | << 1 $) the effective coupling between
two fermions is weak and matter properties can be calculated analytically.
The ground-state energy of normal (i.e. non-superfluid) matter in this regime
was calculated by Lee and Yang in 1957: \cite{Lee:1957}
\begin{equation}
\frac{E}{E_{FG}}  =  1 + \frac{10}{9\pi}k_Fa + \frac{4}{21\pi^2} \left
( 11 - 2 \ln2 \right ) \left ( k_Fa \right )^2~,
\label{eq:Lee}
\end{equation}
where $E_{FG}$ is the energy of a free Fermi gas at the same density as the
interacting gas.  While this expression ignores the contributions of
superfluidity, these are exponentially small in (1/$k_F a$).  In the next
section we compare these results to QMC calculations for $|k_F a| \geq 1$.

The mean-field BCS approach  celebrated in the present volume
reduces in the weak-coupling limit to:
\begin{equation}
\Delta^0_{BCS}(k_F) = \frac{8}{e^2} \frac{\hbar^2 k_F^2}{2m} \exp\left( \frac{\pi}{2ak_F}\right)~.
\label{eq:weakBCS}
\end{equation}
As was shown in 1961 by Gorkov and
Melik-Barkhudarov, \cite{Gorkov:1961} the BCS result acquires a finite
polarization correction even at weak coupling, yielding a reduced pairing gap:
\begin{equation}
\Delta^0 (k_F) = \frac{1}{(4e)^{1/3}} \frac{8}{e^2} \frac{\hbar^2 k_F^2}{2m} \exp\left( \frac{\pi}{2ak_F}\right)~.
\label{eq:weakGMB}
\end{equation}
Thus, the polarization corrections reduce the mean-field BCS result by a factor of
$1 / (4e)^{1/3} \approx 0.45$. Interestingly, if one treats the polarization
effects at the level of sophistication used in the work of Gorkov and
Melik-Barkhudarov, this factor changes with $k_F a$ \cite{Schulze:2001},
though there is no {\it a priori} reason to expect such an approach to be
valid at stronger coupling ($k_F a$ of order 1 or more). Calculating the
pairing gap in this region has been a difficult task, as can be seen from the
multitude of publications devoted to this subject over the past
twenty years.\cite{Lombardo:2001,Dean:2003,Chen:1993,Wambach:1993,
Schulze:1996,Schwenk:2003,Muether:2005,Fabrocini:2005,Cao:2006,
Margueron:2008,Gandolfi:2008a,Abe:2009,Gandolfi:2009a}

\subsection{Quantum Monte Carlo}
\label{sec:qmc} 

Quantum Monte Carlo simulations typically begin with a local Hamiltonian of the form:
\begin{equation}
{\cal{H}} = \sum\limits_{k = 1}^{N}  ( - \frac{\hbar^2}{2m}\nabla_k^{2} )  + \sum\limits_{i<j} v(r_{ij})
+\sum_{i<j<k} v_{ijk}~.
\end{equation}
where $N$ is the total number of particles. In the case of cold atoms the interaction only
acts between opposite spin pairs and is effectively a contact interaction (often simulated
on the computer using a short-range potential).
The neutron-neutron interaction is more complicated, containing 
one-pion exchange at large distances, intermediate range spin-dependent
attraction dominated by two-pion exchange, and a short-range repulsion.
One popular NN interaction that fits the experimental phase shifts well is
Argonne v18~\cite{Wiringa:1995}.
At nuclear densities the two-body force is combined to a three-body force that
is essential to reproduce the spectrum of light nuclei~\cite{Pieper:2001}.
At very low densities, though, as found in neutron star
crusts or the exterior of neutron-rich nuclei, the scattering length ($a = -18.5$ fm)
and effective range ($r_e = 2.7$ fm ) are most crucial to the physical properties of the system. The presence
of a short-range repulsive core is important primarily in that it prevents collapse
to a higher-density state.

Schematically, Green's Function Monte Carlo
projects out the lowest-energy eigenstate
$\Psi_{0}$ from a trial (variational) wave function $\Psi_{V}$ 
by treating the Schr\"{o}dinger equation as a diffusion equation in imaginary
time $\tau$ and evolving the variational wave function up to large $\tau$.
The ground state is arrived at using:
\begin{eqnarray}
\Psi_0 & = & \exp [ - ( H - E_T ) \tau ] \Psi_V  \\ \nonumber
& = & \prod \exp [ - ( H - E_T ) \Delta \tau ] \Psi_V,
\end{eqnarray}
evaluated as a branching random walk.  The short-time propagator is sometimes
taken from a Trotter-Suzuki approximation, but the exact two-nucleon
propagator can also be employed.
The ground-state energy $E_0$ can be obtained from:
\begin{equation}
E_0 = \frac{ \langle \Psi_V | H | \Psi_0 \rangle}{ \langle \Psi_V | \Psi_0 \rangle}
= \frac{ \langle \Psi_0 | H | \Psi_0 \rangle}{ \langle \Psi_0 | \Psi_0 \rangle}.
\end{equation}

In the case of more complicated interactions, one can use either the
nuclear Green's Function Monte Carlo method\cite{Pieper:2001} or the Auxiliary Field Diffusion 
Monte Carlo approach. \cite{Schmidt:1999} Schematically, the latter 
method reduces the spin-isospin dependence of the interaction 
operators from quadratic to linear by means of 
the Hubbard-Stratonovich transformation. 
Given a generic operator $\hat O$ and a parameter $\lambda$:
\begin{equation}
\label{eq:hs}
e^{-\frac{1}{2}\lambda \hat O^2}=\frac{1}{\sqrt{2\pi}}\int dx e^{-\frac{x^2}{2}+\sqrt{-\lambda}x\hat O} \,.
\end{equation}
As a result, the Monte Carlo sampling is no longer limited to the coordinate
space positions of the particles, but now extends to sample auxiliary fields.
Thus, the AFDMC algorithm limits the exponential growth of the
spin-isospin states, and recovers the ground state in polynomial time.

In these Quantum Monte Carlo superfluid simulations the 
trial wave function was taken to be of the Jastrow-BCS form with fixed
particle number: \cite{Carlson:2003}
\begin{equation}
\Psi_V = \prod\limits_{i \neq j} f_P(r_{ij}) \prod\limits_{i' \neq j'} f_P(r_{i'j'}) \prod\limits_{i,j'} f(r_{ij'})  {\cal A} [ \prod_{i<j'} \phi (r_{ij'}) ]
\end{equation}
and periodic boundary conditions. The primed (unprimed) indices correspond to
spin-up (spin-down) neutrons. The pairing function $\phi (r)$ is a sum over
the momenta compatible with the periodic boundary conditions. In the BCS
theory the pairing function is:
\begin{equation}
\phi(r) =\sum\limits_{\bf n} \frac{v_{{\bf k}_{\bf n}}}{u_{{\bf k}_{\bf n}}} e^{i{\bf k}_{\bf n}\cdot{\bf r}} =\sum_{\bf n} \alpha_n e^{i{\bf k}_{\bf n}\cdot{\bf r}} \,.
\label{eq:pairf}
\end{equation}
The Jastrow part is usually taken from a
lowest-order-constrained-variational method \cite{Pandharipande:1973}
calculation described by a Schr\"{o}dinger-like equation. The
fixed-node approximation guarantees that the result for one set of
pairing function parameters will be an upper bound to the
true ground-state energy of the system. Variational results with the pairing function
alone reproduce BCS calculations with finite particle-number projection.
The parameters are optimized in the full
QMC calculation, providing the best possible nodal surface, in the sense of
lowest fixed-node energy, with the given form of trial function.  
Comparisons of fixed-node results\cite{Forbes:2011} in the cold atom system
with recent AFQMC results~\cite{Carlson:2011} indicate that the fixed node calculations are
quite accurate.

In the case of closed-shell neutron drops the antisymmetric part of the trial wave function has the form
\begin{equation}
\psi(R,S) = \Big[\sum Det\{\phi_\alpha(\vec r_i,s_i)\}\Big]_{J,M} \,,
\end{equation}
where $\alpha=\{n,j,m_j\}$ is the set of quantum numbers of single-particle
orbitals, and the summation of more determinants is done in order
to have a trial wave function that is an eigenstate of $J^2$ and $M$.
The single-particle basis is given by
\begin{eqnarray}
\phi_\alpha(\vec r,s)=\Phi_{n,j}(r)\left[Y_{l,m_l}(\hat r)\xi_{s,m_s}(s)\right]_{j,m_j} \,,
\end{eqnarray}
The radial components $\Phi_{n,j}$ are obtained solving the Hartree-Fock
problem with a Skyrme force and their width is variationally
optimized, $Y_{l,m_l}$ are spherical harmonics and $\xi_{s,m_s}$ are
spinors in the usual up-down base.  
It is particularly important at low densities (low oscillator frequencies) to incorporate explicit
BCS correlations in the trial wave function, essentially using single
particle orbitals in the above equation in a form like Eq.~\ref{eq:pairf}.

\section{Infinite Matter Results}
\label{sec:canm}

\begin{figure}[t]
\begin{center}
\includegraphics[width=0.8\textwidth]{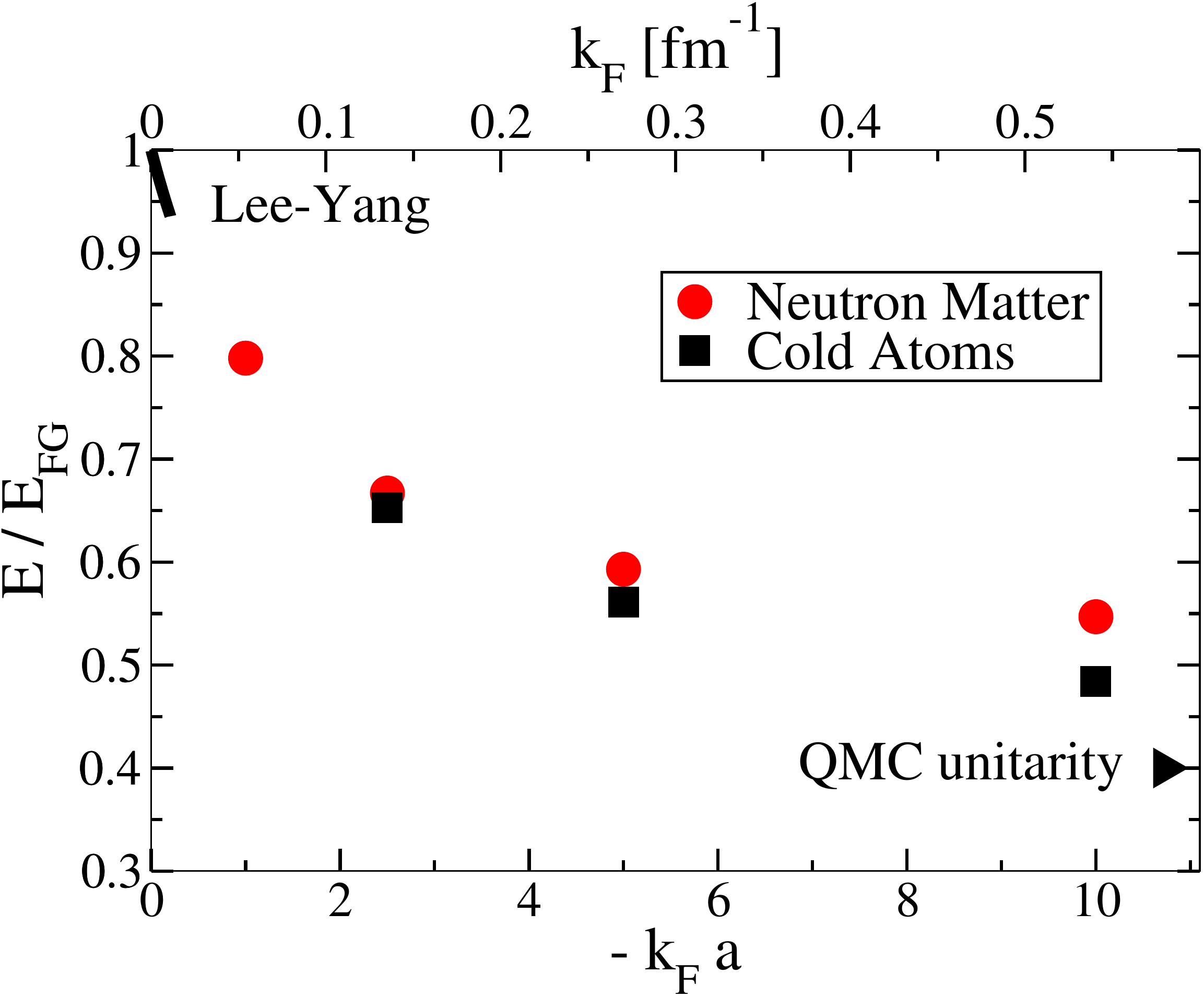}
\caption{Quantum Monte Carlo equation of state for cold atoms (squares) and neutron matter (circles). 
Also shown are the analytic expansion of the ground-state
energy of a normal fluid (line) and an early Quantum Monte Carlo result at unitarity (arrow).}
\label{fig:eos_compare}
\end{center}
\end{figure}

The $T=0$ equations of state for homogeneous cold atoms 
and neutron matter of Ref. \citenum{Gezerlis:2008} are compared in Fig. \ref{fig:eos_compare}.
The horizontal axis is $k_F a$, with the equivalent Fermi momentum $k_F$ for neutron matter shown
along the top. The vertical axis is the ratio of the ground-state energy to the free Fermi gas energy $(E_{FG})$ at the
same density; as discussed in the previous section (Eq. (\ref{eq:Lee})), 
it must go to one at very low densities and decrease as the density increases and
the interactions become important.
The curve at lower densities shows the analytical result by Lee and Yang mentioned previously.
The QMC results shown in this figure seem to agree with the trend implied by the Lee-Yang result. 
The neutron matter and cold atom equations of state are very similar even for densities where the effective range is
comparable to the interparticle spacing.
Near $k_F a = -10$ the energy per particle is not too far from QMC 
calculations and measurements of the ratio $\xi$ of the unitary gas energy 
to $E_{FG}$ shown as an arrow on the right (it corresponds to $k_F a = \infty$); 
previous calculations give $\xi \approx 0.4$, more recent results in Refs.~\citenum{Forbes:2011,Carlson:2011} are slightly lower).
At larger densities the cold-atom and neutron matter results start to diverge, due to: i) the neutron finite effective range, and ii) 
the fact that the neutron results also incorporate a simple attempt to include the $S=1$, $M_S=0$ pairs.
When the density is very low, the $s$-wave contribution is dominant so cold atoms and neutron matter agree very well.
As the density increases, the effective range, as well as higher order terms in momentum and higher partial waves,
become important.

\begin{figure}[t]
\begin{center}
\includegraphics[width=0.8\textwidth]{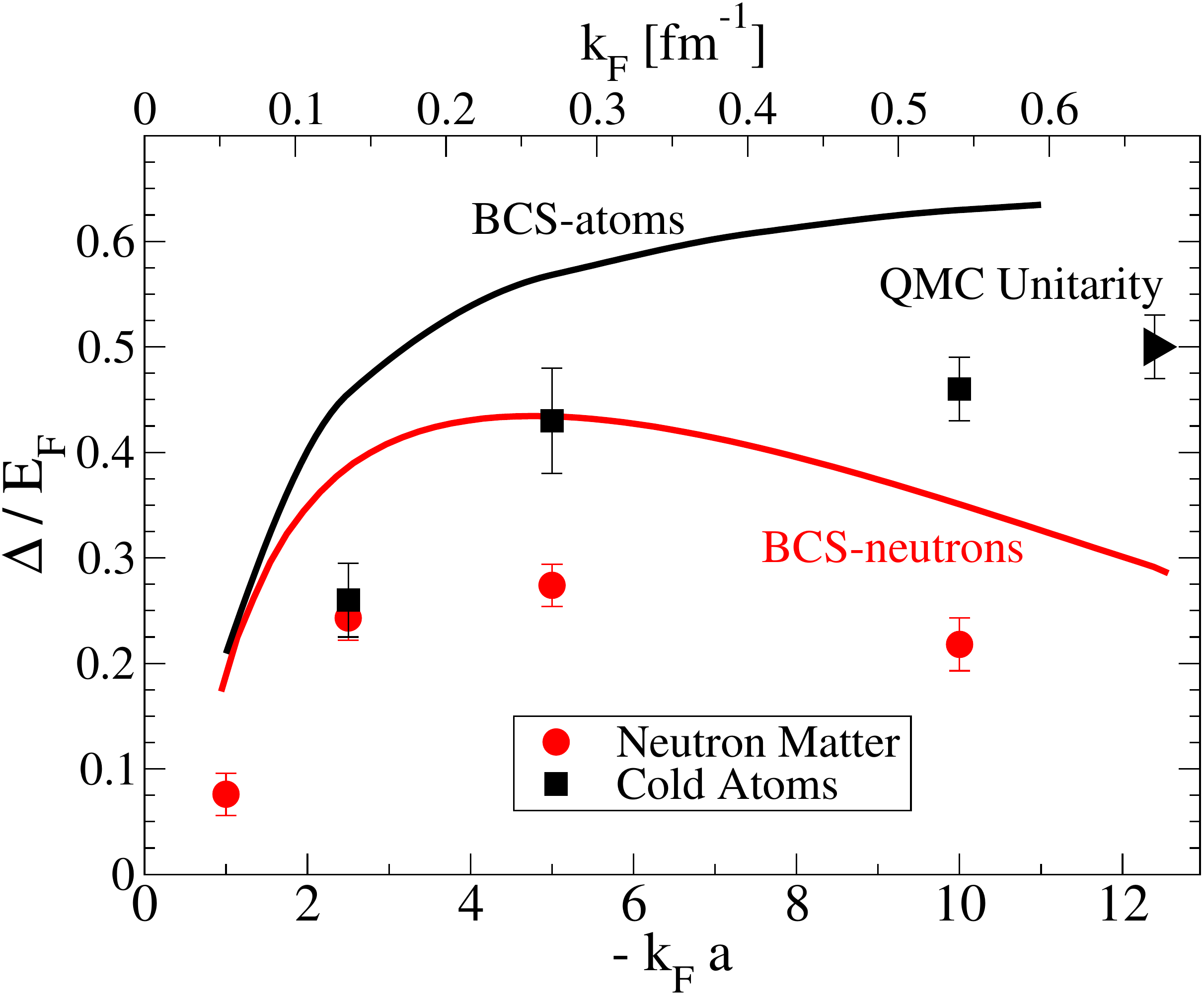}
\caption{Superfluid pairing gaps versus $k_F a$ for cold atoms ($r_e \approx 0$) 
and neutron matter ($ |r_e / a| \approx 0.15$).
BCS (solid lines) and QMC results (points) are shown.}
\label{fig:AFDMClikegap}
\end{center}
\end{figure}

The pairing gap at T=0 is calculated using the odd-even staggering formula:
\begin{equation}
\Delta = \big [ E(N+1) - \frac{1}{2} \left [ E(N)+E(N+2) \right ] \big] (-1)^{N}~,
\label{eq:staggerer}
\end{equation}
where $N$ is the number of particles. 
In Fig. \ref{fig:AFDMClikegap} we plot the gap as a function of $k_F a$
for both cold atoms and neutron matter, taken from Ref. \citenum{Gezerlis:2008}. 
BCS calculations are shown as solid
lines, and QMC results are shown as points with error bars. QMC pairing gaps
are shown from calculations of $N=66-68$ particles.
For very weak coupling,
$ - k_F a << 1$, the pairing gap is expected to be reduced from the BCS
value by the polarization corrections calculated by Gorkov and
Melik-Barkhudarov, $\Delta/ \Delta_{BCS} = (1/4e)^{1/3}$, 
as mentioned in the previous section (Eq. (\ref{eq:weakGMB})). 
The QMC calculations at the lowest density, $k_F a = -1$, are roughly
consistent with this reduction from the BCS value.
At slightly larger yet still small densities,
where $-k_F a = {\cal O} (1)$ but $k_F r_e << 1$ for neutron matter, 
one would expect the pairing gap to be similar for cold atoms and neutron 
matter. The results at $k_F a = -2.5$, where $k_F r_e \approx 0.35$, support 
this expectation. Beyond that density the effective range becomes important 
and both the BCS and the QMC results are significantly reduced in relation to cold atoms where $r_e \approx 0$.

In cold atoms, the suppression from BCS is reduced as the density increases, with a
smoothly increasing fraction of the BCS results as we move from the BCS to the
BEC regime.  At unitarity the measured pairing gaps
\cite{Shin:2008,Carlson:2008,Schirotzek:2008} are 0.45(0.05) of the Fermi
energy, for a ratio $\Delta/\Delta_{BCS} \approx 0.65$,  in agreement with
predictions by QMC methods.\cite{Carlson:2003,Carlson:2005,Gezerlis:2008} 
In the BEC regime where two fermions are tightly bound, the BCS and
QMC values would both give a gap of half the binding energy of the pair.
In
neutron matter, the finite range of the potential reduces $\Delta/E_F$
as the density increases.  We find a ratio $\Delta/\Delta_{BCS}$ that
increases slightly from $|k_F a| = 1$ to $2.5$, but then remains roughly
constant up to $|k_F a| = 10$.

In Refs. \citenum{Gandolfi:2009a,Gezerlis:2010} new QMC values for neutron matter 
were compared to selected previous
results\cite{Chen:1993,Wambach:1993,
Schulze:1996,Schwenk:2003,Fabrocini:2005,Cao:2006,
Gandolfi:2008a,Abe:2009}.
The results of the QMC calculations are much larger than most diagrammatic
approaches. As these approaches assume a well-defined
Fermi surface or calculate polarization corrections based on single-particle
excitations it is not clear how well they can describe neutron matter in the
strongly paired regime, or the similar pairing found in cold atoms. 
Finally, the QMC results seem to qualitatively agree (at least for the lowest densities
considered) with a determinantal Quantum Monte Carlo lattice calculation
\cite{Abe:2009}.

\begin{figure}[t]
\begin{center}
\includegraphics[width=0.9\textwidth]{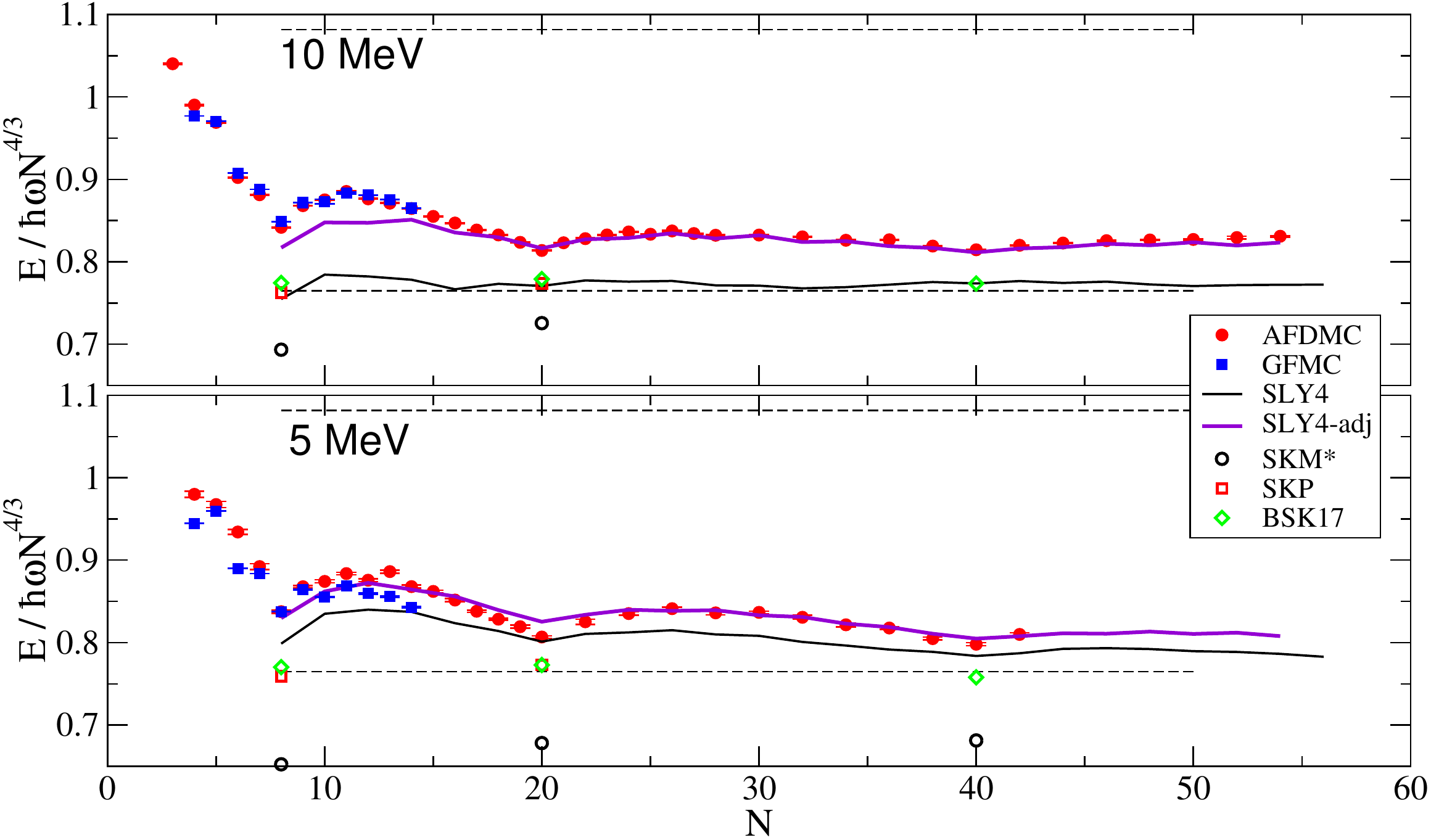}
\caption{Energy of neutrons confined in a harmonic well 
with frequency $\omega=5$ and $10$ MeV. Solid points are ab-initio
calculations computed using AFDMC (red circles) and GFMC (blue squares). 
Open symbols and solid black lines represent results given by selected
Skyrme forces. The violet solid lines show results obtained with the
adjusted SLy4.
}
\label{fig:ene_ndrop}
\end{center}
\end{figure}

\section{Neutron drops}
\label{sec:neutrondrops}

A neutron drop is an idealized system where neutrons, interacting via
realistic nuclear forces, are confined in external potentials.
They provide a simple model to study neutron-rich nuclei. In
Refs.~\citenum{Gandolfi:2006,Gandolfi:2008} the neutron-rich isotopes of
oxygen and calcium have been studied as  neutrons confined in external
fields, and this model provided good results for both bound and excited
states. In addition, these models  can be used to constrain
density functionals in the large isospin-asymmetry region.  
{\it Ab initio} study of neutrons confined in different geometries showed that
some Skyrme forces, typically fitted to nuclei with small isospin-asymmetry,
are not accurate when dealing with pure neutron systems~\cite{Gandolfi:2011}.
The energies computed using QMC  show that Skyrme forces
tend to  overbind confined neutrons, the main contribution to the difference
being the poorly constrained gradient term. In Fig.~\ref{fig:ene_ndrop}
we plot the energy of confined neutrons computed using QMC techniques,
compared with selected  Skyrme forces.

\begin{figure}[t]
\begin{center}
\includegraphics[width=0.9\textwidth]{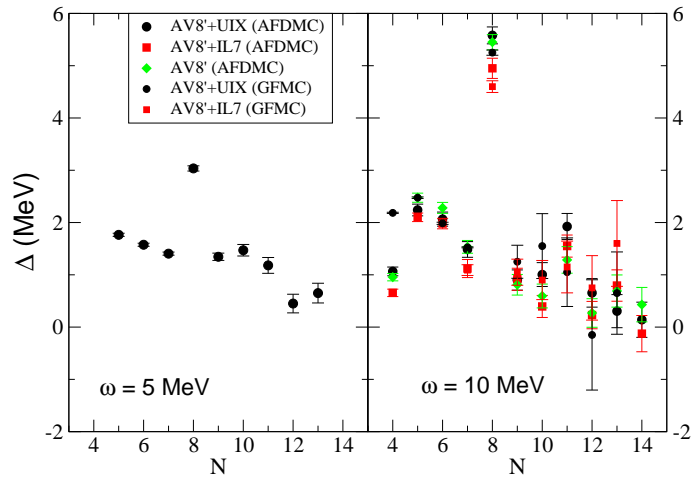}
\caption{Pairing gap of neutrons confined in a harmonic well 
with frequencies $\omega= 5$ (left) and $10$ (right) MeV. Results were arrived at
using different Hamiltonians with only a two-body
force, and including different three-body forces are presented.
}
\label{fig:gap_ndrop}
\end{center}
\end{figure}

The harmonic oscillator potential introduces another element into the pairing
of neutrons, namely the mean field or the spacing of single-particle levels in
the neutron drop. By studying the odd-even staggering of neutron drops with different
frequencies, it is possible to examine the pairing as it evolves from small finite systems
to infinite matter.  The drops we have considered to date have harmonic oscillator frequencies
of 5 and 10 MeV. In both these cases the pairing is much smaller than the single-particle
spacing, so the BCS wave function is largely that occuring in the lowest shell of the drop.
The GFMC calculations have included this pairing into the trial wave function, and have been
used to examine both trap frequencies.  At a frequency of 10 MeV, the pairing is reduced particularly
for the larger drops. Early results are shown in Fig.~\ref{fig:gap_ndrop}
where the gap of confined neutrons is calculated using different Hamiltonians~\cite{Gandolfi:2011,Carlson:2012}.
The dominant physics is provided by the NN interaction, with modest changes introduced by the three-nucleon
interaction. Further studies of these systems are underway.

\section{Summary and Future Work}
\label{sec:future}

In summary, this review has examined the application of Quantum Monte Carlo
methods to strongly-paired matter, including homogeneous and trapped systems of atoms
or neutrons. For infinite matter, the calculated equation 
of state and pairing gap match smoothly with the known 
analytic results at low densities, and provide important constraints in 
the strong-coupling regime at large $k_F a$.  
The low-density equation of state can help constrain 
Skyrme mean-field models of finite nuclei. The pairing gap for 
low-density neutron matter is relevant to Skyrme-Hartree-Fock-Bogoliubov 
calculations \cite{Chamel:2008} of neutron-rich nuclei and to 
neutron-star physics, since it is expected to influence the 
behavior of the crust. \cite{Page:2012} Similarly, results for
neutron drops can also be used to constrain Skyrme and other
energy-density functional approaches.
Furthermore, the newly determined value of the gap 
implies that a new mechanism that makes 
use of superfluid phonons is competitive to the heat conduction 
by electrons in magnetized neutron stars.\cite{Aguilera:2009}  
Another consequence of the gap magnitude is related both to
neutron-star observations and heavy-nuclei phenomenology:
polarized neutron matter may be plausible within the
context of magnetars, and has recently been attacked 
using Quantum Monte Carlo. \cite{Gezerlis:2011,Gezerlis:2012}

Microscopic many-body simulations will undoubtedly continue
to straddle the divide between atomic and nuclear physics.
Such simulations started with two species and equal populations,
soon thereafter moving to spin-polarization. Further examination of 
the evolution of pairing from small to large systems is an intriguing area of study,
including both cold atoms and neutron matter.
For example, one avenue of future research is related to optical lattice experiments with cold atoms:
to first approximation these are equivalent to periodic external potentials. In the nuclear case, an external potential
would allow us to study the static response of neutron matter and would also facilitate the understanding of the impact
on neutron pairing of the ion lattice that exists in a neutron star crust.
Such microscopic results for the static response could provide further constraints on energy-density functionals used
to describe the crust of neutron stars.

\vspace{1cm}

{\bf Acknowledgments} 
This work was supported by 
DOE Grant Nos. DE-FG02-97ER41014 and DE-AC52-06NA25396.
The computations shown were performed at the National Energy
Research Scientific Computing Center (NERSC) and through
Los Alamos Supercomputing.

\bibliographystyle{ws-rv-van}
\bibliography{ws-rv-sample}

\printindex                         
\end{document}